\documentclass[12pt]{article}\usepackage[hyperfootnotes=false]{hyperref}
\usepackage{epsfig}
\usepackage{float}
\usepackage{dsfont}

\usepackage{bbold}

\usepackage{caption}
\usepackage{subcaption}

\usepackage{amsmath}
\usepackage{amssymb}
\usepackage{graphicx}
\newlength{\abstractwidth}
\renewcommand{\thefootnote}{\fnsymbol{footnote}}
\renewcommand{\thanks}[1]{\footnote{#1}} 
\newcommand{\starttext}{
\setcounter{footnote}{0}
\renewcommand{\thefootnote}{\arabic{footnote}}}

\newcommand{\be}{\begin{equation}}
\newcommand{\bea}{\begin{eqnarray}}
\newcommand{\eea}{\end{eqnarray}}
\newcommand{\beq}{\begin{equation}}
\newcommand{\ee}{\end{equation}}

\def\simleq{\; \raise0.3ex\hbox{$<$\kern-0.75em
\raise-1.1ex\hbox{$\sim$}}\; }
\def\simgeq{\; \raise0.3ex\hbox{$>$\kern-0.75em
\raise-1.1ex\hbox{$\sim$}}\; }

\def\bi{\begin{itemize}}
\def\ei{\end{itemize}}

\def\sc{\setcounter{equation}{0}}

\def\bn{\bigskip \noindent}

\makeatletter
\g@addto@macro\normalsize{%
  \setlength\abovedisplayskip{10pt}
  \setlength\belowdisplayskip{20pt}
  \setlength\abovedisplayshortskip{10pt}
  \setlength\belowdisplayshortskip{20pt}
}
\makeatother

\usepackage{color}

\renewcommand{\title}[1]{\vbox{\center\LARGE{#1}}\vspace{5mm}}
\renewcommand{\author}[1]{\vbox{\center#1}\vspace{5mm}}
\newcommand{\address}[1]{\vbox{\center\em#1}}

\begin{document}
  
\begin{titlepage}

\rightline{}
\bigskip
\bigskip\bigskip\bigskip\bigskip
\bigskip

\centerline{\Large \bf {Dear Qubitzers,}}

\bn

\centerline{\Large \bf {GR=QM.}}

\bigskip

\bigskip
\begin{center}

\author{Leonard Susskind}

\address{Stanford Institute for Theoretical Physics and Department of Physics, \\
Stanford University, Stanford, CA 94305-4060, USA}

\end{center}

\begin{center}
\bf     \rm

\bigskip

\end{center}

\begin{abstract}

These are some thoughts contained in a letter to colleagues, about the close relation between gravity and quantum mechanics, and also about the possibility of seeing quantum gravity in a lab equipped with quantum computers. I expect this will become feasible sometime in the next decade or two.

\medskip
\noindent
\end{abstract}

\end{titlepage}

\starttext \baselineskip=17.63pt \setcounter{footnote}{0}

\vfill\eject


\vfill\eject


\bn
7/28/2007

\bn
Dear Qubitzers,

\bn
GR=QM? Well why not? Some of us already accept ER=EPR \cite{Maldacena:2013xja}, so why not follow it to its logical conclusion?

\bn

It is said that general relativity and quantum mechanics are separate subjects that don't fit together comfortably. There is a tension, even a contradiction between them---or so one often hears.
 I take exception to this view.   I think that  exactly the opposite is true. It may be too strong to say that gravity and quantum mechanics are exactly the same thing, but   those of us who are paying attention, may already  sense that the two are inseparable, and  that neither makes sense without the other.

Two things make me think so. The first is  ER=EPR, the equivalence between  quantum entanglement and spatial connectivity. In its strongest form ER=EPR holds not only for black holes but for any entangled systems---even empty space%
\footnote{Empty space can be divided by Rindler horizons, so that the two sides are entangled   \cite{VanRaamsdonk:2010pw}\cite{Ryu:2006bv}.}.
  If the entanglement between two spatial regions is somehow broken, the regions become disconnected \cite{VanRaamsdonk:2010pw}; and conversely, if regions are entangled, they must be connected \cite{Maldacena:2013xja}.  The most basic property of space---its connectivity---is due to the most quantum property of quantum mechanics: entanglement. Personally I think ER=EPR is more than half way to GR=QM.

The second   has to do with the dynamics of space, in particular its tendency to expand. One sees this in cosmology, but also behind the horizons of black holes.   The expansion is thought to be connected with the  tendency  of quantum states to become increasingly complex. Adam Brown and I called this tendency  \it the second law of quantum complexity\rm \ \cite{Brown:2017jil}.  If one pushes these ideas  to their logical limits, quantum entanglement of any kind implies the existence of hidden Einstein-Rosen bridges which  have a strong tendency to grow, even in situations which one naively would think have nothing to do with gravity.

 To summarize this viewpoint in a short slogan:

\bn
\it  Where there is quantum mechanics, there is also gravity.\rm

\bn
I suggest that this is true in a very strong sense;  even for systems that are deep into the non-relativistic  range of parameters---the range in which the Newton constant is negligibly small, and  and the speed of light  is much larger than any laboratory velocity. 
This may sound like a  flight of  fantasy, but I believe it is an inevitable consequence of things we already accept.

\bn

Let's suppose that a laboratory exists, containing a large spherical shell made of some more or less ordinary material, that is well described by non-relativistic quantum mechanics. That's all there is in the lab except for a few simple devices like strain gauges, squids to measure magnetic field, and a light hammer to tap on the shell. One other thing: Alice and Bob to do some experiments. Obviously---you say---quantum gravity is completely irrelevant in this lab: a perfect counter example to my claim that ``where there is quantum mechanics there is also gravity." Equally obviously, I don't agree.

 The   shell has been engineered to be  at a quantum critical point, where the excitations  are described by a  conformal field theory  having    a holographic  bulk%
\footnote{By bulk I mean the  AdS-like geometry dual to the CFT. The space in the laboratory will just be called \it the lab\rm. The bulk should not be confused with the ordinary interior of the shell which is part of the lab.  
\ 

Because a  real shell has a finite number of atoms the CFT is a cutoff version, which means that the bulk geometry terminates on some physical boundary finitely far from the center of the bulk.}
 dual. 
Assume that the  shell has a  signal  velocity much less than the laboratory speed of light, and that the gravitational constant in the lab is so small that gravitational effects on the shell are negligible. Experiments on the shell would be limited only by the laws of quantum mechanics, and not by the speed of light or by  gravitational back reaction. 

You can probably see where this is going, but you may be tempted to respond:
 ``You are cleverly simulating a system that has a bulk dual, but it's just a simulation, not  real quantum gravity."  Again, I disagree.

I  argue that the   bulk  with  its gravitons,  black holes, and  bulk observers is just as real as the laboratory itself. It can be probed, entered, measured, and the results communicated to observers in the lab---even in the limit that the laboratory  laws of physics  become arbitrarily close to non-relativistic quantum mechanics.

\bn

From the holographic AdS/CFT  correspondence we may assume that observers, perhaps with human-like cognitive abilities, are possible in the bulk. Can a laboratory observer  confirm the presence of  such a bulk observer  by doing experiments on the shell? Current theory seems to say yes. By tapping on the shell the laboratory observer can perturb the CFT, exciting low dimension operators. In the gravity dual these perturbations  source bulk gravitational waves and other light fields. An observer floating at the center of AdS would detect these signals with a bulk LIGO detector. In that way the laboratory observer can communicate with the bulk observer.

Similarly the bulk observer can send messages to the lab. Gravitational  signals emitted in  the bulk  will reach the boundary, and be detected as disturbances of the shell by means of  strain gauges, squids,  or other devices. The bulk and laboratory observers can communicate with one another and even carry on a conversation. Responding to questions from the lab, the bulk observer may report the existence of  massless spin-2 particles.  In that way  the lab observer learns that  quantum gravity exists in the bulk.

It is possible that the CFT describing the shell does not  have  a weakly coupled standard gravitational dual. If the central charge (number of independent CFT field degrees of freedom) is not large, the entire radius of the AdS universe would not be large in bulk Planck units. Or if the CFT coupling were too weak other microscopic bulk scales (such as the size of strings) would be large; a single string could fill the bulk space. This would not mean that there is no bulk; it would mean the bulk laws are in some sense messy.  Theories with weakly coupled  gravity duals are special limiting cases, but bulk duals for  the other cases should not be ruled out. Pushing this to its logical conclusion: any quantum system has a gravitational dual; even a pair of entangled electron spins has a tiny quantum wormhole through which  a single qubit can be teleported.

Returning to shells that have  weakly coupled gravitational duals, what if the bulk of such a shell has no observer? In principle the lab observer can create one by applying appropriate perturbations to the shell. In fact there is nothing to prevent her from merging her own quantum state  with the shell and entering into the bulk.

\bn

\begin{figure}[H]
\begin{center}
\includegraphics[scale=.3]{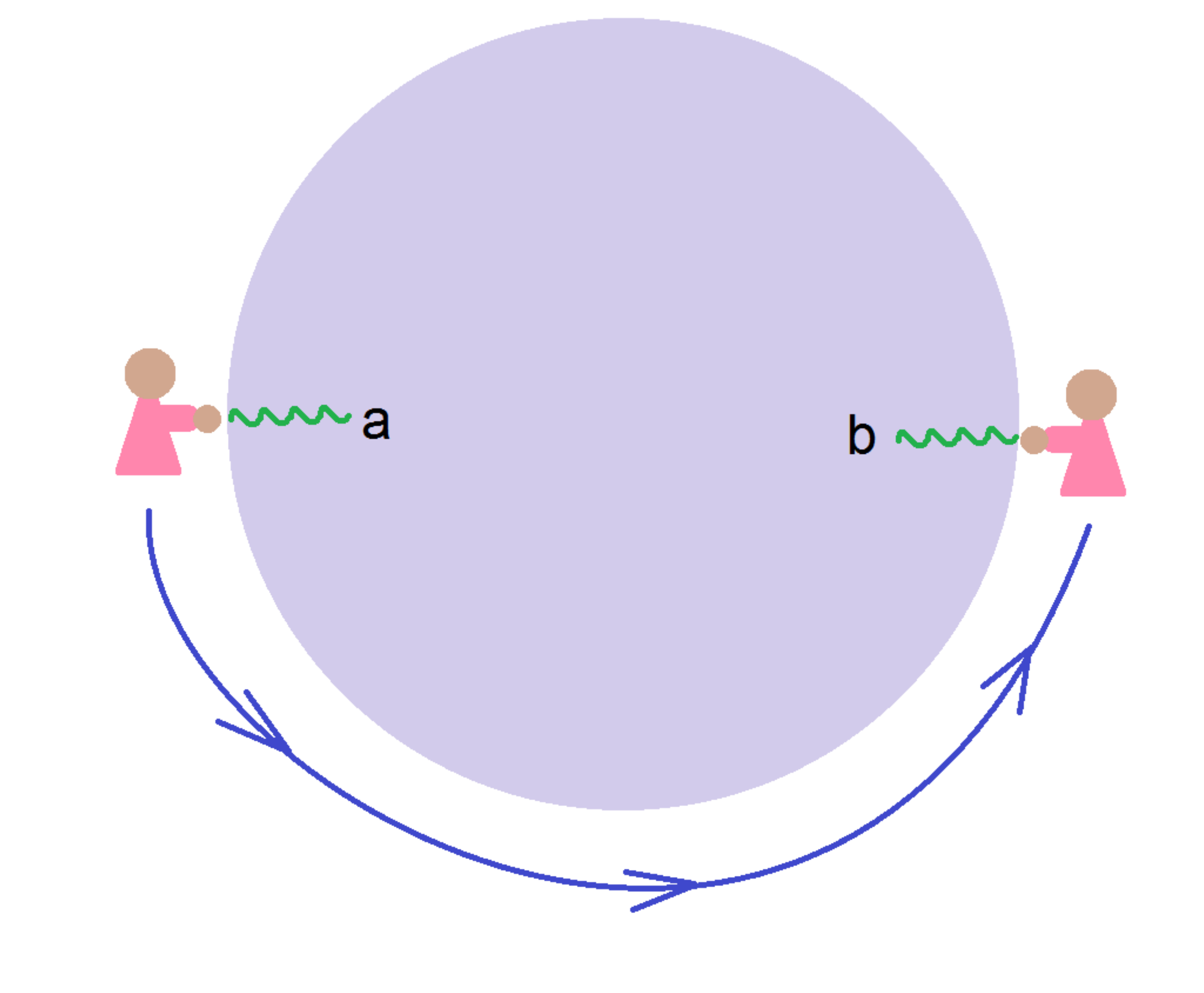}
\caption{}
\label{quick}
\end{center}
\end{figure}

There is an apparent contradiction between this view and bulk locality%
\footnote{I thank Ying Zhao for pointing this out.}.
%
%
%
Figure  \ref{quick} illustrates the point.
\bn
In the bulk a signal originates at point $\bf a $  and travels outward toward the AdS boundary.  Alice plucks the signal out of the bulk and   quickly runs to the opposite side of the shell  where she  re-inserts the signal. The signal then propagates from the boundary to point $\bf b$, the whole  trip taking less time than  for light, traveling through the bulk,  to go from $\bf a$ to $\bf b$.

All this is true, but it is not a contradiction with bulk locality. No one in the bulk sees a signal move past them with a local velocity faster than light. What is true is that the boundary conditions (at the AdS boundary) are not of the usual  reflecting type. For example non-local boundary conditions  in AdS/CFT  can be induced by double-trace bilocal operators added to the CFT action. The bulk observers may be astonished by how fast signals can propagate across  the boundary, but they do not see a bulk violation of the speed of light. 

\bn

What about bulk black holes? Given some energy the bulk observer can create a small black hole and report  its properties to the lab. Alternately the lab observer can heat the shell and create a large black hole.  But the most interesting questions about black holes might remain unanswered; namely what  happens beyond the horizon? Since no signal can get to the boundary from behind the  horizon, there would seem to be no way for the lab observer to confirm the existence of a black hole interior.

I believe this  conclusion is too pessimistic. 
Let's suppose that the lab contained two well-separated identical shells, entangled in the thermofield-double state. The bulk dual, sometimes called an eternal black hole \cite{Maldacena:2001kr},  is really two entangled black holes. According to our usual understanding of gauge-gravity duality the bulk description contains an Einstein-Rosen bridge (ERB) connecting the two entangled black holes.

On the other hand, since the laboratory is part of a world governed by quantum mechanics and by gravity, we may apply the principles of quantum gravity to the lab. In particular we may assume ER=EPR, \bf \underline{not} \rm  for the  bulk theory, but for the laboratory itself.
It would imply that the two entangled shells are  connected by an ERB of some sort. Are these two the same---the ERB of the bulk dual and the ERB of the entangled laboratory shells? 

I believe they are.

Assume that the two entangled shells are controlled by laboratory technicians Alice and Bob.   In principle Alice should be able to merge a laboratory system---call it Tom-the-teleportee---with her shell, so that effectively Tom is injected into the bulk. In the bulk description Tom may fall into Alice's black hole and crash into the singularity.  Bob may also inject a system---Tina---into his shell, and if conditions are right, Tom and Tina can meet in the wormhole  before they are destroyed at the singularity. But because they  can't send a signal to the boundary from behind the horizon, Tom and Tina can't tell Alice and Bob about their experiences. 
One might come to the conclusion that the world behind the horizon is a figment of the imagination, with no operational significance or possibility of being falsified.

But is this conclusion warranted? I don't think so. To see why, recall that an entangled system  can be used to mediate quantum teleportation---in this case the teleportation of Tom from Alice to Bob.  
Carrying this out would require Alice to transfer a certain number  of classical bits  from  her shell to Bob's.  The transfer of classical information takes place through the laboratory space%
\footnote{Since the  speed  of signal propagation  in the shell may be much smaller  than the speed of light, from the bulk point of view the classical exchange takes essentially no time. },
not through the wormhole, but the classical information  is completely random and uncorrelated with  Tom's quantum state. In fact by the monogamy of entanglement, there can be no trace of Tom's quantum state in the lab.  

Then how did Tom's qubits  get from Alice to Bob if they did not pass through the lab? The usual quantum theorist's answer is that quantum information is non-locally distributed in the entangled state; it doesn't make sense to ask where it is localized. But ER=EPR suggests another answer \cite{Susskind:2014yaa}\cite{Susskind:2016jjb}: Quantum teleportation is \it teleportation through the wormhole. \rm
In other words teleported information is transferred between entangled systems by passing through the Einstein-Rosen bridge connecting them.

This conclusion has recently gained significant credibility due to the  work of Gao, Jafferis, and Wall \cite{Gao:2016bin}, and the subsequent followup by Maldacena, Stanford, and Yang \cite{Maldacena:2017axo}. What these authors show is that the same conditions that allow quantum teleportation, have a dynamical back-reaction on the wormhole that renders it traversable.

 What makes the protocol for  teleportation through the wormhole  special---what makes it, \it through the wormhole\rm---is that once Tom has merged with Alice's shell, enough time is allowed to elapse so that  his information becomes  scrambled with the horizon degrees of freedom. This takes place before the rest of the protocol is executed \cite{Susskind:2017nto}. To put it another way, the first step of the protocol is to allow Tom 
  to fall through the horizon of Alice's black hole. 

It is especially interesting  that if Tom encounters  objects during his passage from Alice's side to Bob's side, his experiences may be recorded in his memory%
\footnote{This is especially clear in \cite{Gao:2016bin}
 and \cite{Maldacena:2017axo} }. This would allow him to report the conditions in the wormhole to  laboratory observers. 
 
Quantum teleportation through the wormhole is a real  game-changer; it provides a direct way to observe the interior geometry of a wormhole. One can  no longer  claim that life behind the horizon is unphysical, meaningless, unobservable, or scientifically unfalsifiable.

\bn

Can laboratory experiments of this type be carried out? I don't see why not. Instead of  shells supporting conformal field theories, a more practical alternative might be quantum computers simulating the CFTs. Entangling two identical quantum computers into a thermofield double state should be feasible. 
To teleport a genuine sentient Tom through  the  wormhole would require an enormous number of qubits, but with a few hundred logical qubits one can teleport a register, composed of  say ten qubits---enough for a primitive memory.   By varying the initial entangled state one can vary the environment in the wormhole. In turn these variations will  couple to the register and  be recorded, later to be   communicated to the lab
\footnote{Note that quantum teleportation of a single qubit through a channel of a single bell pair is already an experimental reality. ER=EPR allows it to be interpreted as teleportation through a Planckian wormhole.}.

The operations needed for this kind of teleportation are fairly complex (in the computational sense) \cite{Susskind:2017nto} and are therefore difficult, but I don't see anything forbidding them once quantum computers become available.

\bn

One thing that I want to emphasize is that there is no need for 
Planckian energy in the laboratory in order to exhibit these quantum gravity effects. The Planck scale in the bulk is not related to the Planck scale in the lab, but rather to central charge of the CFT. 

\bn

Let me dispel one concern---a counter argument to the claim that black holes dual to thermally excited non-relativistic shells are real. The argument goes as follows:
According to Lloyd \cite{Lloyd} and to my own papers with Brown, Roberts, Swingle, and Zhao 
\cite{Brown:2015bva}\cite{Brown:2015lvg}, black holes are the fastest possible computers. Because the speed of propagation in non-relativistic  shells  is much smaller than the laboratory speed of light, the \it simulated \rm black holes are nowhere near saturating Lloyd's bound on computational speed. The laboratory observer can easily tell the difference between the slow simulated black holes and real black holes.

The error in this argument is the classic logical fallacy: \it All politicians are liars. Therefore Pinocchio is a politician. \rm The correct statement about black holes is that all things that saturate the Lloyd bound are black holes---not that all black holes saturate the Lloyd bound. In   \cite{Brown:2015lvg} an example of a black hole encased in a static ``Dyson sphere"  showed that black holes do not generally saturate the Lloyd bound.

\bn

\bn

\begin{figure}[H]
\begin{center}
\includegraphics[scale=.2]{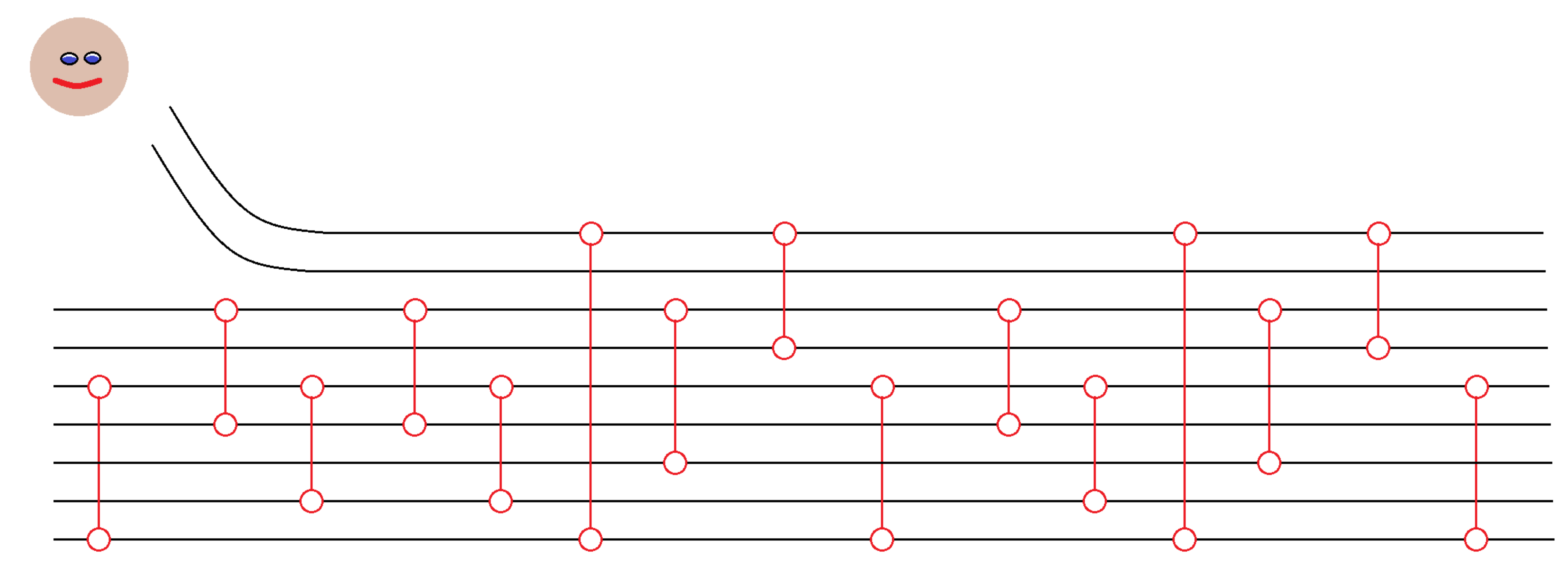}
\caption{}
\label{moon}
\end{center}
\end{figure}

If there is anything new here
\footnote{The ideas in this letter are not particularly original.  On several occasions I've discussed similar things with Juan Maldacena, Joe Polchinski, Don Marolf, and Aron Wall, among others. Related concepts appear in  \cite{Maldacena:2001kr}\cite{Heemskerk:2012mn}\cite{Marolf:2012xe}\cite{Maldacena:2013xja}\cite{Gao:2016bin}\cite{Maldacena:2017axo} and probably a great number of other places.}
 it is the idea that information may pass from a laboratory environment to the degrees of freedom of a  physical realization of a CFT, thereby bridging the gap between the lab and the bulk. One can enter the bulk, observe it, and go back to the lab. From the laboratory point of view this gives operational  meaning to the bulk and to whatever  it contains. Figure \ref{moon} is a cartoonish attempt to illustrate how an observer can be merged with a quantum-circuit computer. The first step would be to act on the observer with a unitary operation that transfers his quantum state to a set of qubits. The qubits can then be combined with the circuit. An appropriate protocol, similar to quantum teleportation,  can move the observer back to the lab at a later time \cite{Maldacena:2017axo}.

\bn

\bn

Now let's turn to quantum complexity and how it governs the growth of space \cite{Susskind:2014rva}\cite{Stanford:2014jda}\cite{Susskind:2015toa}\cite{Brown:2015bva}\cite{Brown:2015lvg}. Einstein-Rosen bridges are not static objects.  They grow with time in a manner that loosely resembles the cosmological expansion of space. This  growth of geometry has a quantum dual description, namely  the statistical growth of quantum computational complexity. 

Why do ERBs grow? From the viewpoint of classical general relativity, the Einstein field equations in combination with positive energy conditions predict such growth. Classically the growth continues forever, but quantum mechanically the finite number of quantum states limits the time of growth to be exponential in the entropy.

On the other side of the duality the growth of complexity reflects a general quantum-statistical law that parallels the second law of thermodynamics. The second law of thermodynamics is about the increase of entropy as systems tend toward thermal equilibrium. The second law of complexity \cite{Brown:2017jil} is about the growth of quantum complexity, which eventually leads to complexity equilibrium  after 
an exponential time. 

The second laws, both of entropy and  complexity, are very general. They apply to black holes, shells of matter, and quantum circuits. The match between the growth of complexity for generic quantum systems, and the expansion of Einstein-Rosen bridges, is remarkably detailed   \cite{Stanford:2014jda}\cite{Roberts:2014isa}. It extends to all kinds of out-of-equilibrium situations including the gravitational back  reaction to violent shock waves. The pattern is so general that laboratory observers, monitoring the growth of complexity,  will see  behaviors that are completely consistent with the relativistic evolution of Einstein-Rosen bridges.

But complexity is a notoriously difficult thing to measure, even if one could do repeated experiments and collect statistics.  There are more direct things that would indicate a growth of the wormhole.  One more or less standard way would be to measure correlation functions between fields at opposite ends of the wormhole. The magnitude of the correlation is a direct measure of the distance between the ends. 

If the two systems (shells, quantum computers, black holes) were unentangled, the relevant distance governing correlations would be the ordinary exterior distance, i.e., the distance outside the horizon separating  two black holes, or the laboratory distance separating the two shells or computers. In general any correlation will fall to zero with increased separation. On the other hand if the systems are entangled the correlation will not decrease with laboratory separation. No matter how far the shells are separated from one another, the correlations between them will behave as if there is a short connection whose length is independent of the exterior distance. In other words the correlation functions behave as if there is a wormhole connecting the systems.

Because wormholes grow with time, the correlation functions should decrease  in a characteristic way%
\footnote{Ordinary local field correlations decrease until noisy fluctuations dominate the falling signal. This happens after a polynomial time \cite{Maldacena:2001kr}. There are various ways to filter out the noise and construct correlators that continue to decrease for an exponential time.}.
If the dynamics of the laboratory systems is generic the evolution of correlations between Alice's and Bob's systems will also be time dependent, despite the fact that both systems are in thermal equilibrium. One can determine the correlation
functions  by making measurements on both systems and collecting statistics. It would not be hard to show that the correlations decrease with time the same way as they would for ``real" black holes with growing ERBs. 

The lab observers could interpret this in two ways. They could say that the correlations decrease because of time-dependence of the phases in the wave function of the combined system, or they could be bold and say the wormhole that mediates the correlations  grows with time. The results would be the same.  

 One might ask if there is a more detailed relation between gravitational dynamics and the properties of the evolution of complexity---a relation which goes beyond the overall expansion of space? The answer is probably yes, and to support that claim I would point to the close relation between complexity of quantum states and the Einstein-Hilbert action of certain regions of bulk space-time \cite{Brown:2015bva}\cite{Brown:2015lvg}. It seems possible that gravitational dynamics can be recovered from the generic behavior of quantum complexity.


\bn

A skeptic may argue that all of this  can be explained without ever invoking bulk gravity or wormholes; plain old quantum mechanics and some condensed matter physics or quantum circuitry is enough. This is absolutely  true, but I think it misses the point:  Theories with gravity are always holographic and require a lower dimensional  non-gravitational description%
\footnote{Of the two,  the bulk description is often much simpler than the strongly coupled holographic description.}. 
This does not mean the bulk world  is not real. 

\bn

Finally, in what way is gravity  special? Suppose instead of a shell of matter with a CFT description, we construct a large  block of matter engineered to have the \it standard model \rm (without gravity) as its excitations;  or  a quantum computer composed of a three-dimensional array of interacting qubits. 
Is the world in the block or  the computer  real? Sure it is; the block and its excitations are certainly real, and if the standard model was well simulated it may support observers who could communicate with laboratory observers. 

But there are some big differences.
In the case of the block, the bulk space really is the three-dimensional volume of the block. It exists in ordinary laboratory space. 
By contrast, in the  case of CFT-supporting shells,  something much more subtle  is at work. The bulk is not part of ordinary space: it is not the shell: it is not the hollow space inside the shell. These are all part of the lab.  
The bulk space  is a pure manifestation of entanglement and complexity. 

\bn

All of this leads me to the conjecture that where there is quantum mechanics there is also gravity, or more succinctly, GR=QM. I'll also bet that once quantum computers become cheap, easy to build, and easy to control,  experiments similar to the ones I've described will become common, and the idea of quantum gravity in the lab will seem much less crazy. In fact such experiments are already on the drawing board \cite{Swingle:2016var} \cite{Yao:2016ayk}.

Best regards,

Lenny

\section*{P.S.}

\begin{figure}[H]
\begin{center}
\includegraphics[scale=.35]{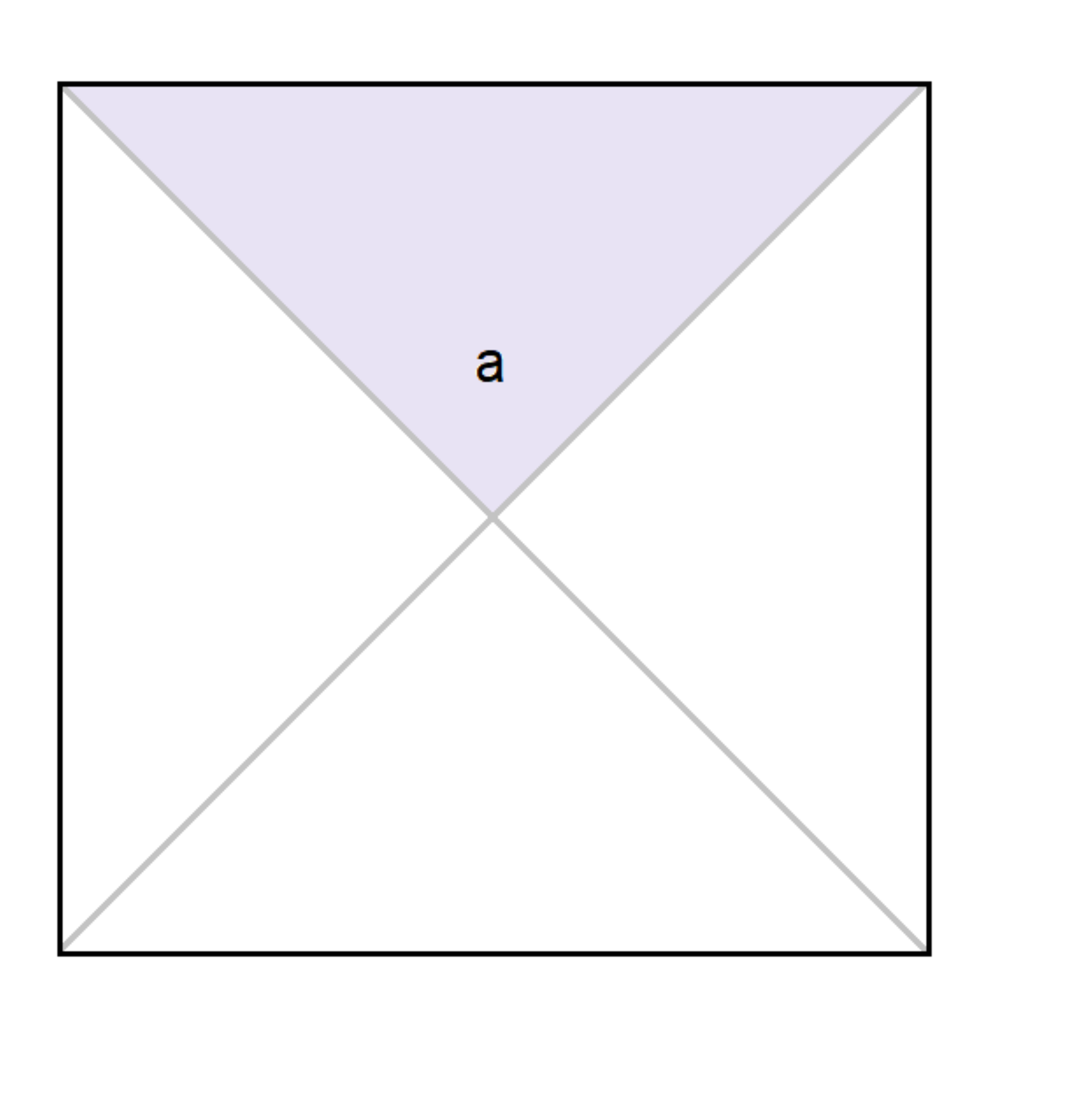}
\caption{The event $\bf a$ is behind the horizon and cannot be seen from the boundary.}
\label{behind}
\end{center}
\end{figure}

 Hrant Gharibyan has pointed out that 
the question of whether teleportation can reveal events behind the horizon is a bit tricky. Consider the event labeled $\bf a$ in figure \ref{behind}. The region behind the future horizon is shaded and the event $\bf a$ is behind the horizon.

\begin{figure}[H]
\begin{center}
\includegraphics[scale=.3]{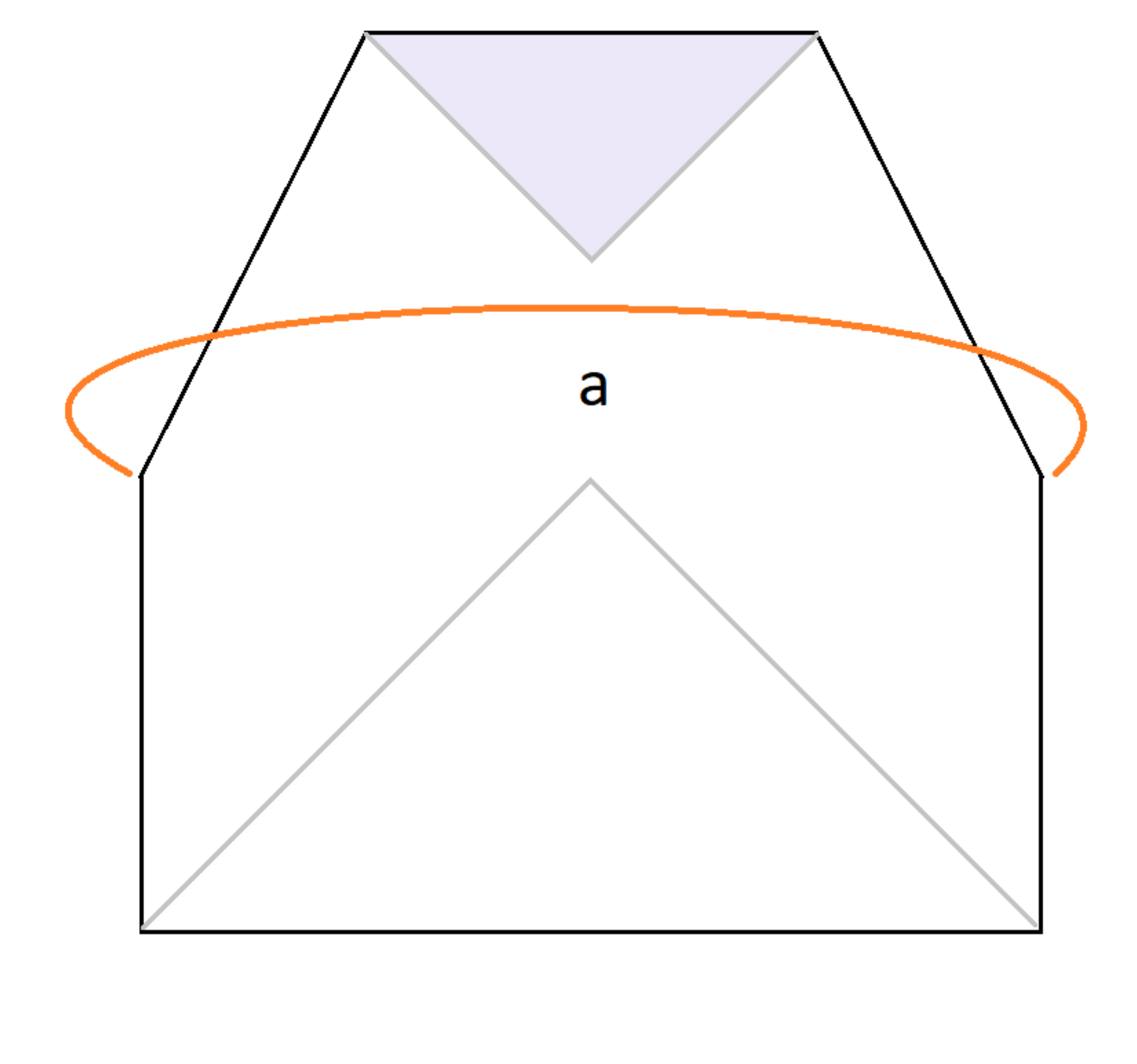}
\caption{The protocol of 
\cite{Gao:2016bin}\cite{Maldacena:2017axo} makes the wormhole traversable so that $\bf a$ becomes visible from the boundaries. The orange curve represents the classical information sent from Alice to Bob, required to carry out the protocol.}
\label{not-behind}
\end{center}
\end{figure}

Now let us apply the protocol of 
\cite{Gao:2016bin}\cite{Maldacena:2017axo} in order to make the ERB traversable. This is shown in figure \ref{not-behind}.

\bn
The event $a$ has been exposed so that Tom can witness it as he passes through the ERB. In this sense the protocol has made visible an event behind the horizon.

But in another sense the horizon has not been breached; it has been moved to the new shaded region.  The rule is that teleportation allows us to see what would be behind the horizon had we not applied teleportation. But we can also say the the very attempt to see behind the horizon moves the horizon to where we cannot see behind it.

\section*{Acknowledgements}

I am  grateful for many discussions over the years which shaped these views. I especially recall discussions with Juan Maldacena, Joe Polchinski, Don Marolf, Aaron Wall, Steve Giddings, Douglas Stanford, Dan Harlow, Ying Zhao, Hrant Gharibyan, and Ben Freivogel. I've probably forgotten many others and for that I apologize. 

Support came through NSF Award Number 1316699.

\sc
\appendix


\begin{thebibliography}{99}

\bibitem{Maldacena:2013xja} 
  J.~Maldacena and L.~Susskind,
  ``Cool horizons for entangled black holes,''
  Fortsch.\ Phys.\  {\bf 61}, 781 (2013)
  doi:10.1002/prop.201300020
  [arXiv:1306.0533 [hep-th]].

\bibitem{VanRaamsdonk:2010pw} 
  M.~Van Raamsdonk,
  ``Building up spacetime with quantum entanglement,''
  Gen.\ Rel.\ Grav.\  {\bf 42}, 2323 (2010)
  [Int.\ J.\ Mod.\ Phys.\ D {\bf 19}, 2429 (2010)]
  doi:10.1007/s10714-010-1034-0, 10.1142/S0218271810018529
  [arXiv:1005.3035 [hep-th]].

\bibitem{Ryu:2006bv} 
  S.~Ryu and T.~Takayanagi,
  ``Holographic derivation of entanglement entropy from AdS/CFT,''
  Phys.\ Rev.\ Lett.\  {\bf 96}, 181602 (2006)
  doi:10.1103/PhysRevLett.96.181602
  [hep-th/0603001].


\bibitem{Brown:2017jil} 
  A.~R.~Brown and L.~Susskind,
  ``The Second Law of Quantum Complexity,''
  arXiv:1701.01107 [hep-th].


\bibitem{Maldacena:2001kr} 
  J.~M.~Maldacena,
 ``Eternal black holes in anti-de Sitter,''
  JHEP {\bf 0304}, 021 (2003)
  doi:10.1088/1126-6708/2003/04/021
  [hep-th/0106112].


\bibitem{Susskind:2014yaa} 
  L.~Susskind,
  ``ER=EPR, GHZ, and the consistency of quantum measurements,''
  Fortsch.\ Phys.\  {\bf 64}, 72 (2016)
  doi:10.1002/prop.201500094
  [arXiv:1412.8483 [hep-th]].


\bibitem{Susskind:2016jjb} 
  L.~Susskind,
  ``Copenhagen vs Everett, Teleportation, and ER=EPR,''
  Fortsch.\ Phys.\  {\bf 64}, no. 6-7, 551 (2016)
  doi:10.1002/prop.201600036
  [arXiv:1604.02589 [hep-th]].


\bibitem{Gao:2016bin} 
  P.~Gao, D.~L.~Jafferis and A.~Wall,
  ``Traversable Wormholes via a Double Trace Deformation,''
  arXiv:1608.05687 [hep-th].


\bibitem{Maldacena:2017axo} 
  J.~Maldacena, D.~Stanford and Z.~Yang,
  ``Diving into traversable wormholes,''
  Fortsch.\ Phys.\  {\bf 65}, no. 5, 1700034 (2017)
  doi:10.1002/prop.201700034
  [arXiv:1704.05333 [hep-th]].


\bibitem{Susskind:2017nto} 
  L.~Susskind and Y.~Zhao,
  ``Teleportation Through the Wormhole,''
  arXiv:1707.04354 [hep-th].


\bibitem{Heemskerk:2012mn} 
  I.~Heemskerk, D.~Marolf, J.~Polchinski and J.~Sully,
 ``Bulk and Transhorizon Measurements in AdS/CFT,''
  JHEP {\bf 1210}, 165 (2012)
  doi:10.1007/JHEP10(2012)165
  [arXiv:1201.3664 [hep-th]].


\bibitem{Marolf:2012xe} 
  D.~Marolf and A.~C.~Wall,
  ``Eternal Black Holes and Superselection in AdS/CFT,''
  Class.\ Quant.\ Grav.\  {\bf 30}, 025001 (2013)
  doi:10.1088/0264-9381/30/2/025001
  [arXiv:1210.3590 [hep-th]].





\bibitem{Susskind:2014rva} 
  L.~Susskind,
  ``Computational Complexity and Black Hole Horizons,''
  Fortsch.\ Phys.\  {\bf 64}, 24 (2016)
  doi:10.1002/prop.201500092
  [arXiv:1403.5695 [hep-th], arXiv:1402.5674 [hep-th]].


\bibitem{Stanford:2014jda} 
  D.~Stanford and L.~Susskind,
  ``Complexity and Shock Wave Geometries,''
  Phys.\ Rev.\ D {\bf 90}, no. 12, 126007 (2014)
  doi:10.1103/PhysRevD.90.126007
  [arXiv:1406.2678 [hep-th]].


\bibitem{Susskind:2015toa} 
  L.~Susskind,
  ``The Typical-State Paradox: Diagnosing Horizons with Complexity,''
  Fortsch.\ Phys.\  {\bf 64}, 84 (2016)
  doi:10.1002/prop.201500091
  [arXiv:1507.02287 [hep-th]].


\bibitem{Brown:2015bva} 
  A.~R.~Brown, D.~A.~Roberts, L.~Susskind, B.~Swingle and Y.~Zhao,
  ``Holographic Complexity Equals Bulk Action?,''
  Phys.\ Rev.\ Lett.\  {\bf 116}, no. 19, 191301 (2016)
  doi:10.1103/PhysRevLett.116.191301
  [arXiv:1509.07876 [hep-th]].
  
\bibitem{Brown:2015lvg} 
  A.~R.~Brown, D.~A.~Roberts, L.~Susskind, B.~Swingle and Y.~Zhao,
  ``Complexity, action, and black holes,''
  Phys.\ Rev.\ D {\bf 93}, no. 8, 086006 (2016)
  doi:10.1103/PhysRevD.93.086006
  [arXiv:1512.04993 [hep-th]].


\bibitem{Roberts:2014isa} 
  D.~A.~Roberts, D.~Stanford and L.~Susskind,
  ``Localized shocks,''
  JHEP {\bf 1503}, 051 (2015)
  doi:10.1007/JHEP03(2015)051
  [arXiv:1409.8180 [hep-th]].

\bibitem{Lloyd}
S. Lloyd, ``Ultimate physical limits to computation," Nature 406 (2000),
no. 6799 1047–1054.


\bibitem{Swingle:2016var} 
  B.~Swingle, G.~Bentsen, M.~Schleier-Smith and P.~Hayden,
  ``Measuring the scrambling of quantum information,''
  Phys.\ Rev.\ A {\bf 94}, no. 4, 040302 (2016)
  doi:10.1103/PhysRevA.94.040302
  [arXiv:1602.06271 [quant-ph]].

\bibitem{Yao:2016ayk} 
  N.~Y.~Yao, F.~Grusdt, B.~Swingle, M.~D.~Lukin, D.~M.~Stamper-Kurn, J.~E.~Moore and E.~A.~Demler,
  ``Interferometric Approach to Probing Fast Scrambling,''
  arXiv:1607.01801 [quant-ph].

\end{thebibliography}
\end{document}